\documentclass[final]{svjour3}
\usepackage{graphicx}
\usepackage{rotating}
\usepackage{amssymb}
\usepackage{mathptmx}
\usepackage[numbers]{natbib}
\usepackage{bm}

\makeatletter
\journalname{Journal of Low Temperature Physics}

\def\lesssim{\ \raise.3ex\hbox{$<$}\kern-0.8em\lower.7ex\hbox{$\sim$}\ }
\def\gesim{\ \raise.3ex\hbox{$>$}\kern-0.8em\lower.7ex\hbox{$\sim$}\ }

\bibpunct{}{}{,}{s}{}{,}

\begin{document}
\newcommand{\hdblarrow}{H\makebox[0.9ex][l]{$\downdownarrows$}-}
\title{Non-equilibrium properties of a pumped-decaying Bose-condensed electron-hole gas in the BCS-BEC crossover region}
\author{R. Hanai${}^{1}$ \and P. B. Littlewood${}^{2,3}$ \and Y. Ohashi${}^{1}$}\institute{
1: Department of Physics, Keio University, 3-14-1, Hiyoshi, Kohoku-ku, Yokohama 223-8522, Japan\\
Tel.: +81-45-566-1454 \\ Fax: +81-45-566-1672 \\
\email{rhanai@rk.phys.keio.ac.jp}\\
2: Physical Sciences and Engineering, Argonne National Laboratory, Argonne, Illinois 60439, USA\\
3: University of Chicago, James Frank Institute, Chicago, Illinois 60637, USA}

\date{\today}
\maketitle
\begin{abstract}
We theoretically investigate a Bose-condensed exciton gas out of equilibrium. Within the framework of the combined BCS-Leggett strong-coupling theory with the non-equilibrium Keldysh formalism, we show how the Bose-Einstein condensation (BEC) of excitons is suppressed to eventually disappear, when the system is in the non-equilibrium steady state. The supply of electrons and holes from the bath is shown to induce quasi-particle excitations, leading to the partial occupation of the upper branch of Bogoliubov single-particle excitation spectrum. We also discuss how this quasi-particle induction is related to the suppression of exciton BEC, as well as the stability of the steady state.


\end{abstract}
\section{Introduction}
Since the prediction of an exciton Bose-Einstein condensation (BEC) in a semiconductor\cite{Blatt1962,Keldysh1968}, this electron-hole pair condensate has attracted much attention as an analogous phenomenon to metallic superconductivity, where electron-electron Cooper-pairs play the central role. Although the exciton BEC has not been realized yet, recent experiments have explored sub-Kelvin temperatures\cite{Stolz2012,Yoshioka2013,Alloing2014}. Thus, the realization of an exciton BEC is very promising. 
\par
Once this Fermi condensate is realized, it is expected that one can examine various physical properties of this system from the weak-coupling regime to the strong-coupling limit, by adjusting the exciton density. At a glance, this advantage is similar to the case of a superfluid Fermi atomic gas, where the interaction between Fermi atoms is also tunable by adjusting the threshold energy of a Feshbach resonance\cite{Chin}. In the latter, the so-called BCS-BEC crossover has been realized\cite{Jin,Ketterle}, where the character of superfluidity continuously changes from the weak-coupling BCS type to the BEC of tightly bound molecules, with increasing the strength of a pairing interaction\cite{Eagles1969,Leggett1980,Comte1982}. However, while the cold atom system is usually in the equilibrium state, an exciton gas is essentially in the non-equilibrium state, because one always needs to continue supplying electrons and holes to the system, in order to compensate the decay of excitons into photons\cite{Moskalenko}. Thus, the realization of an exciton BEC would provide a unique opportunity to examine the BCS-BEC crossover phenomena in the non-equilibrium case. Since non-equilibrium properties of a Fermi condensate has recently been discussed in various systems, such as an exciton-polariton gas in a microcavity\cite{Kasprzak2006}, as well as an ultracold Fermi gas\cite{Falkenau2011,Mahnke2015}, an exciton BEC would also contribute to the study in these fields.
\par
In this paper, we investigate non-equilibrium properties of an exciton condensate. We employ a model for an electron-hole gas, that has attractive interactions between the species to promote pairing, and also decay to a vacuum and pumping from a bath of free fermions. The tunneling of particles to the vacuum effectively describes the decay of excitons in this model. This leakage is compensated by the supply of particles from the bath, leading to the steady state. In this model, we examine strong-coupling effects within the BCS-Leggett theory at zero bath temperature, $T_{\rm b}=0$. Effects of the non-equilibrium steady state are also taken into account by using the Keldysh Green's function\cite{Rammer,Szymanska2006,Yamaguchi2012}. In the non-equilibrium steady state, we examine how the exciton BEC is suppressed, to eventually disappear. As a signature of this suppression, we show that partial occupation of the upper branch of Bogoliubov single-particle excitations occurs. Throughout this paper, we take $\hbar=k_{\rm B}=1$, and system volume is set to unity, for simplicity.
\par
\par
\section{Non-equilibrium BCS-Leggett theory in the presence of pumping and decay}
\par
We consider a model electron-hole gas in the BEC state, described by the Hamiltonian,
\begin{equation}
H=H_{\rm s}+H_{\rm env}+H_{\rm t},
\label{H}
\end{equation}
where,
\begin{eqnarray}
H_{\rm s}&=&
\sum_{\bm p} \Psi^\dagger_{\bm p} 
\left[
\varepsilon_{\bm p} \tau_3 - \Delta(t) \tau_+ - \Delta^*(t) \tau_- 
\right]\Psi_{\bm p}
-U\sum_{\bm q}\rho_{\bm q}^+\rho_{-{\bm q}}^-,
\label{Hs}
\\
H_{\rm env}&=&
\sum_{\bm p}\Phi_{\bm p}^{{\rm b}\dagger}
\varepsilon_{\bm p}^{\rm b} \tau_3 \Phi_{\bm p}^{\rm b}
+
\sum_{\bm p}\Phi_{\bm p}^{{\rm v}\dagger}
\varepsilon_{\bm p}^{\rm v} \tau_3 \Phi_{\bm p}^{\rm v},
\label{Henv}
\\
H_{\rm t} &=& \sum_{\lambda={\rm b,v}} \sum_{{\bm p},{\bm q}}\sum_i
\left[\Gamma_\lambda \Phi^{\lambda\dagger}_{\bm q} \tau_3 \Psi_{\bm p}e^{i{\bm p}\cdot{\bm r}_i}e^{-i{\bm q}\cdot{\bm R}_i} + {\rm h.c.}
\right].
\label{Ht}
\end{eqnarray}
The electron-hole gas in the exciton-BEC phase is described by $H_{\rm s}$ in Eq. (\ref{Hs}), where $\Psi_{\bm p}=(a_{\bm p,{\rm e}}, a^\dagger_{-\bm p,{\rm h}})^{\rm T}$ is a Nambu field, consisting the electron annihilation operator ($a_{\bm p,{\rm e}}$) and the hole creation operator ($a^\dagger_{-\bm p,{\rm h}}$). These particles are assumed to have the same mass $m$, as well as the same kinetic energy $\varepsilon_{\bm p}=p^2/(2m)$. $\tau_i$ ($i=1,2,3$) are Pauli matrices acting on electron-hole space, and $\rho_{\bm q}^\pm=\sum_{\bm p}\Psi_{{\bm p}+{\bm q}}\tau_\pm\Psi_{\bm p}$, where $\tau_\pm=[\tau_1\pm i\tau_2]/2$. The attractive interaction between an electron and a hole is modeled by a contact interaction with the coupling constant $-U~(<0)$ (although the real interaction is, of course, the long-range Coulomb interaction). For simplicity, we ignore the repulsive interaction between electrons, as well as the interaction between holes. The exciton-BEC state is characterized by the order parameter $\Delta(t)=U\sum_{\bm k} \langle a_{-\bm k,{\rm h}}(t)a_{\bm k,{\rm e}}(t)\rangle$. 
\par
The model exciton-BEC gas is coupled to a bath and a vacuum described by the first and second term in $H_{\rm env}$ in Eq. (\ref{Henv}), respectively. In. Eq. (\ref{Henv}), the Nambu field $\Phi^{\rm b(v)}_{\bm p}=(c^{\rm b(v)}_{\bm p,{\rm e}}, c^{{\rm b(v)}\dagger}_{-\bm p,{\rm h}})^{\rm T}$ consists of the electron annihilation operator $c^{\rm b(v)}_{\bm p,{\rm e}}$ and the hole creation operator $c^{{\rm b(v)}\dagger}_{-\bm p,{\rm h}}$ in the bath (vacuum). Electrons and holes in the bath (vacuum) are assumed to have the same kinetic energy $\varepsilon_{\bm p}^{\rm b(v)}$, for simplicity.
\par
The momentum-independent transfer matrix element $\Gamma_{\rm b}$ ($\Gamma_{\rm v}$) in Eq. (\ref{Ht}) represents the coupling between the exciton-BEC system and the bath (vacuum). Here, we assume that electrons and holes of the exciton BEC at position ${\bm r}_i$ tunnels to position ${\bm R}_i$ in the bath (vacuum). While the leakage of particles from the exciton-BEC gas into the vacuum effectively describes the decay of excitons, the tunneling from the bath to the exciton-BEC works as the pumping, to compensate the leakage of particles into the vacuum. 
\par
For the time dependence of the BEC order parameter $\Delta(t)$ in the non-equilibrium steady state, we employ the ansatz\cite{Szymanska2006,Yamaguchi2012} 
\begin{equation}
\Delta(t)=\Delta_0 e^{-2i\mu t},
\label{Delta}
\end{equation}
where $\Delta_0$ is taken to be real. In this case, one can formally eliminate the time dependence from the model Hamiltonian in Eq. (\ref{H}) by the gauge transformation $(a_{\bm p},c_{\bm p}^{\rm b},c_{\bm p}^{\rm v})=({\tilde a}_{\bm p},{\tilde c}_{\bm p}^{\rm b},{\tilde c}_{\bm p}^{\rm v})e^{-i\mu t}$. The resulting Hamiltonian has the same form as Eq. (\ref{H}), where $\varepsilon_{\bm p}$, $\varepsilon_{\bm p}^{\rm b,v}$, and $\Delta(t)$, are replaced by $\xi_{\bm p}=\varepsilon_{\bm p}-\mu$, $\xi_{\bm p}^{\rm b,v}=\varepsilon_{\bm p}^{\rm b,v}-\mu$, and $\Delta_0$, respectively.
\par
As usual, we assume that the bath and the vacuum are huge compared to the exciton-BEC system, so that they are still in the thermal equilibrium state, even when they are coupled to the exciton-BEC system. In particular, we consider the case where the bath is at $T_{\rm b}=0$. In this case, the electron and hole distribution in the bath is simply given by the ordinary Fermi distribution function at $T_{\rm b}=0$, $f_{\rm b}(\omega)=\theta(\omega-[\mu_{\rm b}-\mu])$, where $\mu_{\rm b}$ is the Fermi energy in the bath. For the vacuum, since the particles are absent there, we take the vanishing distribution in the vacuum as $f_{\rm v}(\omega)\equiv 0$.
\par
\begin{figure}
\begin{center}
\includegraphics[width=0.62\linewidth,keepaspectratio]{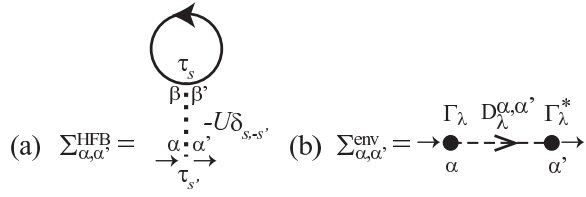}
\end{center}
\caption{(Color online) Diagrammatic expressions for the self-energy. (a) $\Sigma_{\alpha,\alpha'}^{\rm HFB}$. (b) $\Sigma_{\alpha,\alpha'}^{\rm env}$. The solid line and the dashed line represent the Keldysh Green's function in the exciton-BEC system and that in the bath or vacuum, respectively. The dotted line describes the electron-hole interaction $-U$, and solid circle represents the transfer matrix element $\Gamma_{\rm b,v}$ between the exciton-BEC system and the environment (consisting of a bath and a vacuum).
}
\label{fig1}
\end{figure}
\par
\par
\begin{figure}
\begin{center}
\includegraphics[width=1.0\linewidth,keepaspectratio]{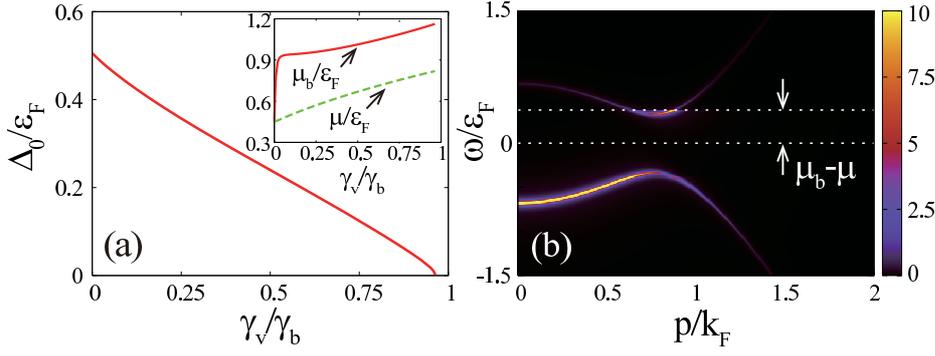}
\end{center}
\caption{(Color online) (a) Exciton-BEC order parameter $\Delta_0$ in the unitarity limit ($(k_{\rm F}a_s)^{-1}=0$) at $T_{\rm b}=0$, as a function of the decay parameter $\gamma_{\rm v}$. We take $\gamma_{\rm b}/\varepsilon_{\rm F}=10^{-2}$, where $\varepsilon_{\rm F}$ is the Fermi energy of a free electron-hole gas. The inset shows $\mu_{\rm b}$ and $\mu$. (b) Intensity of the occupied spectral weight $L(\bm p, \omega)$, normalized by $\varepsilon_{\rm F}^{-1}$. We set $\gamma_{\rm v}/\gamma_{\rm b}=0.3$. The other parameters are the same as those used in panel (a).
}
\label{fig2}
\end{figure}

To systematically examine non-equilibrium effects on the strong-coupling exciton BEC at $T_{\rm b}=0$, it is convenient to reformulate the BCS-Leggett theory using the Keldysh Green's function $G_{\alpha,\alpha'}({\bm p},\omega)$\cite{Rammer}, which obeys the Dyson equation
\begin{eqnarray}
G_{\alpha,\beta}(\bm p,\omega) 
=G^{0}_{\alpha,\alpha'}(\bm p,\omega) + \sum_{\beta,\beta'}
G^0_{\alpha,\beta}(\bm p,\omega) \Sigma_{\beta,\beta'}(\bm p,\omega) 
G_{\beta',\alpha'}(\bm p,\omega), 
\label{Dyson}
\end{eqnarray}
where the lowest-order Keldysh Green's function $G_{\alpha,\alpha'}^0(\bm p,\omega)$ has the form, under the Nambu representation,
\begin{eqnarray}
{\hat G}^0(\bm p,\omega)=
\{G^0 \}_{\alpha,\alpha'}=
\left(
\begin{array}{cc}
G_{\rm R}^0 & G_{\rm K}^0 \\
0 & G_{\rm A}^0
\end{array}
\right)
=
\left(
\begin{array}{cc}
{\displaystyle 1 \over \displaystyle \omega+i\delta-\xi_{\bm p}\tau_3} & -2\pi i\delta(\omega-\xi_{\bm p}\tau_3){\rm sgn}\omega \\
0 & {\displaystyle 1 \over \displaystyle \omega-i\delta-\xi_{\bm p}\tau_3}
\end{array}
\right).
\label{G0}
\end{eqnarray}
In Eq. (\ref{G0}), $\delta$ is an infinitesimally small positive number, and $G_{\rm R}^0$, $G_{\rm A}^0$, and $G_{\rm K}^0$, are the retarded, advanced, and Keldysh components, respectively. The self-energy $\Sigma_{\alpha,\alpha'}(\bm p,\omega)$ in Eq. (\ref{Dyson}) involves effects of the electron-hole interaction ($-U$), as well as the coupling to the bath ($\Gamma_{\rm b}$) and the vacuum ($\Gamma_{\rm v}$). In the BCS-Leggett theory, the former is treated within the Hartree-Fock-Bogoliubov (HFB) approximation, where the HFB self-energy $\Sigma^{\rm HFB}_{\alpha,\alpha'}(\bm p,\omega)$ is diagrammatically described as Fig. \ref{fig1}(a), which gives \cite{Schrieffer}
\begin{equation}
\Sigma_{\alpha,\alpha'}^{\rm HFB}(\bm p, \omega) 
=iU\sum_{\bm p'}\int {d\omega' \over 2\pi}
\sum_{\beta,\beta'}\sum_{s,s'=\pm} 
\eta^{\alpha,\beta}_{\alpha',\beta'} \delta_{s,-s'} 
{\rm Tr}[\tau_s G_{\beta',\beta}(\bm p',\omega')]\tau_{s'}.
\label{SigHFB}
\end{equation}
Here, $\eta ^{\alpha,\beta}_{\alpha',\beta'}=(\delta_{\alpha,\alpha'}\delta_{\beta,-\beta'}+\delta_{\alpha,-\alpha'}\delta_{\beta,\beta'})/2$, where $-\alpha'$ means the opposite component to $\alpha'$. We also include the couplings to the bath ($\Gamma_{\rm b}$) and the vacuum ($\Gamma_{\rm v}$) in the second Born approximation shown diagrammatically in Fig.\ref{fig2}(b). The expression for this self-energy correction $\Sigma^{\rm env}$ is given by, after taking the random average over the tunneling position ${\bm r}_i$, 
\begin{eqnarray}
\Sigma^{\rm env}_{\alpha,\alpha'}({\bm p},\omega) 
=\sum_{\lambda={\rm b,v}}
N_{\rm t}|\Gamma_\lambda|^2\sum_{\bm q}D_\lambda^{\alpha,\alpha'}({\bm q},\omega)
=
\sum_{\lambda={\rm b,v}}
\left(
\begin{array}{cc}
-i\gamma_\lambda & 
-2i\tau_3\gamma_\lambda F_\lambda(\omega\tau_3)
\\
0 & i\gamma_\lambda
\end{array}
\right)_{\alpha,\alpha'}.
\label{Sigenv}
\end{eqnarray}
Here, $D_\lambda^{\alpha,\alpha'}({\bm q},\omega)$ are the non-interacting Keldysh Green's functions in the bath ($\lambda={\rm b}$) and the vacuum ($\lambda={\rm v}$), $F_\lambda=1-2f_\lambda(\omega)$, and
\begin{equation}
\gamma_\lambda=\pi N_{\rm t}\rho_\lambda|\Gamma_\lambda|^2
\label{gamma}
\end{equation}
describes pumping and decay effects by the coupling to the bath ($\lambda={\rm b}$) and vacuum ($\lambda={\rm v}$) ($N_{\rm t}$ is the number of tunneling positions). In obtaining Eq. (\ref{gamma}), the bath and the vacuum are assumed to be white, that is, each part has the energy-independent single-particle density of states $\rho_\lambda$. 
\par
Using the self-energy $\Sigma=\Sigma^{\rm HFB}+\Sigma^{\rm env}$, one obtains the self-consistent equation from $G_{11}(\bm p,\omega)$ in Eq. (\ref{Dyson}), which corresponds to the BCS gap equation in the equilibrium state, as
\begin{eqnarray}
1=U \sum_{\bm p}\int {d\omega \over 2\pi}
{
{\tilde F}(\omega)[\omega+\xi_{\bm p}+i\gamma]
-{\tilde F}(-\omega)[\omega-\xi_{\bm p}-i\gamma]
\over
[(\omega - E_{\bm p})^2 + \gamma^2]
[(\omega + E_{\bm p})^2 + \gamma^2]
},                            
\label{GAP}
\end{eqnarray}
where $\gamma=\gamma_{\rm b}+\gamma_{\rm v}$, and ${\tilde F}(\omega)=\gamma_{\rm b}F_{\rm b}(\omega)+\gamma_{\rm v}F_{\rm v}(\omega)$. $E_{\bm p}=\sqrt{\xi_{\bm p}^2 + \Delta_0^2}$ is the ordinary the Bogoliubov single-particle excitation spectrum. Since Eq. (\ref{GAP}) involves the ultraviolet divergence, as in the ordinary BCS gap equation, we need to renormalize the theory to eliminate this singularity. This is conveniently achieved by measuring the interaction strength in terms of the $s$-wave scattering length $a_s$, which is related to the contact interaction $-U$ as ${4\pi a_s / m}= -U / [1-U\sum_{\bm p}^{p_{\rm c}}(1/ (2\varepsilon_{\bm p}) ) ]. $ In this scale, $(k_{\rm F}a_s)^{-1}\lesssim 0$ and $0\gesim(k_{\rm F}a_s)^{-1}$  represent the weak-coupling side and the strong-coupling side, respectively (where $k_{\rm F}$ the Fermi momentum). $p_c$ is a cutoff momentum.
\par
Following the BCS-Leggett theory in the equilibrium state \cite{Eagles1969,Leggett1980}, we solve the ``gap equation" (\ref{GAP}), together with the equation for the total number $N$ of electrons and holes, which is obtained from $(1,1)$-component (in Nambu space) of the lesser Green's function $G^<=-G_{11}+G_{22}+G_{12}$ as\cite{Rammer}
\begin{eqnarray}
N&=&-2i\sum_{\bm p} \int {d\omega \over 2\pi}
G_{11}^<({\bm p},\omega)
\nonumber
\\
&=&
2\sum_{\bm p} \int {d\omega \over \pi}
{
[(\omega+ \xi_{\bm p})^2+\Delta_0^2+\gamma^2]\gamma_{\rm b}f_{\rm b}(\omega)
+
\Delta_0^2[\gamma_{\rm b}(1- f_{\rm b}(-\omega))+\gamma_{\rm v}]
\over
[(\omega - E_{\bm p})^2 +\gamma^2]
[(\omega + E_{\bm p})^2 +\gamma^2],
},
\label{NUM}
\end{eqnarray}
to self-consistently determine $\Delta_0, \mu_{\rm B}$ and $\mu$, for a given parameter set $(N, a_s,\gamma_{\rm b},\gamma_{\rm v})$. In this regard, we briefly note that the gap equation (\ref{GAP}) is a complex equation, so that the real and imaginary components give two independent equations. Thus, together with the number equation (\ref{NUM}), one may safely determine these three quantities in a consistent manner.
\par
\par
\section{Depairing effect in the non-equilibrium exciton-BEC phase}
\par
Figure \ref{fig2}(a) shows the self-consistent solutions for the coupled equations (\ref{GAP}) and (\ref{NUM}). In the equilibrium case (which is realized when $\gamma_{\rm v}=0$), the inset shows the expected relation $\mu_{\rm b}=\mu$. When this chemical equilibrium condition is satisfied, the imaginary part of the gap equation (\ref{GAP}) identically vanishes (Note that $F_{\rm b}(-\omega)=-F_{\rm b}(\omega)$ when $\mu_{\rm b}=\mu$.), and the real part of this equation is reduced to the ordinary BCS gap equation in the limit, $\gamma_{\rm b}\to 0$. 
\par
When the leakage of electrons and holes into the vacuum is turned on ($\gamma_{\rm v}>0$), the bath must supply particles to the exciton-BEC system so as to compensate this leakage. Because of this, the chemical potential $\mu_{\rm b}$ in the bath rapidly increases when $\gamma_{\rm v}$ is small but finite. (See the region around $\gamma_{\rm v}=0$ in the inset in Fig. \ref{fig2}(a).) We also see in Fig. \ref{fig2}(a) that the magnitude of the exciton-BEC order parameter $\Delta_0$ decreases, which physically means that the BEC phase is suppressed in the non-equilibrium state, even when the particle loss is compensated by the bath. In the unitarity limit shown in Fig. \ref{fig2}(a), the exciton-BEC phase disappears when $\gamma_{\rm v}/\gamma_{\rm b}\ge0.96$.
\par
The decrease of the order parameter $\Delta_0$ in the presence of pumping ($\gamma_{\rm b}>0$) and decay ($\gamma_{\rm v}>0$) implies that the depairing effect occurs in the non-equilibrium exciton-BEC. To confirm this in a simple manner, we consider the occupied spectral weight $L(\bm p, \omega)$, which directly gives information about occupied single-particle states, given by,
\begin{eqnarray}
L(\bm p, \omega)&=&-iG_{11}^{<}(\bm p,\omega)
\nonumber\\
&=&
\left[ 
{\gamma_{\rm b} \over \gamma}f_{\rm b}(\omega)\zeta(\bm p,\omega) 
+
\Bigl[
{\gamma_{\rm b} \over \gamma}[1-f_{\rm b}(-\omega)]
+
{\gamma_{\rm v} \over \gamma}
\Bigr]
[1-\zeta(\bm p,\omega)] 
\right]
A(\bm p,\omega ),
\label{OSW}
\end{eqnarray}
Here, $\zeta(\bm p,\omega )=[(\omega+\xi_{\bm p})^2+\gamma^2]/[(\omega+\xi_{\bm p})^2+\Delta_0^2+\gamma^2]$, and
\begin{eqnarray}
A(\bm p, \omega)=
{1 \over 2\pi}\left[1+{\xi_{\bm p} \over E_{\bm p}}\right]
{\gamma \over (\omega- E_{\bm p})^2+\gamma^2}
+
{1 \over 2\pi}\left[1-{\xi_{\bm p} \over E_{\bm p}}\right]
{\gamma \over (\omega +E_{\bm p})^2+\gamma^2}
\label{SW}
\end{eqnarray}
is the spectral weight. In the equilibrium state ($\gamma_{\rm v}=0$), the occupied spectral weight $L(\bm p, \omega)$ is reduced to (Note that $\mu_{\rm b}=\mu$.)
\begin{equation}
L(\bm p, \omega)=\Theta(-\omega)A({\bm p},\omega),
\label{OSWeq}
\end{equation}
where $\Theta(x)$ is the step function. Equation (\ref{OSWeq}) indicates that the upper Bogoliubov single-particle band ($\omega=E_{\bm p}$) is almost unoccupied, when $\gamma_{\rm b}/\Delta_0\ll 1$. However, in the non-equilibrium state, Fig. \ref{fig2}(b) shows the sizable occupation of this branch, when the ``biased voltage"  $\mu_{\rm b}-\mu$ exceeds the energy gap $\Delta_0$. Since the total number $N$ of electrons and holes remains unchanged in the steady state, the appearance of the Bogoliubov (unpaired) quasi-particles is accompanied by the decrease of the condensate fraction in the exciton-BEC.
\par
The condition $\mu_{\rm b}-\mu\gesim\Delta_0$ is also obtained as the condition to maintain the steady state, when $\gamma_{\rm b}/\Delta_0\ll 1$. In this limiting case, the imaginary part of Eq. (\ref{GAP}) gives
\begin{eqnarray}
\gamma_{\rm b}\sum_{\bm p}
{\gamma \over E_{\bm p}^2+\gamma^2}\Theta(-E_{\bm p}+\mu_{\rm b}-\mu)
=\gamma_{\rm v}\sum_{\bm p}
{\gamma \over E_{\bm p}^2+\gamma^2}.
\label{GAPim}
\end{eqnarray}
To satisfy this, the condition 
\begin{equation}
\mu_{\rm b}-\mu\ge {\rm Min}[E_{\bm p}]=\Delta_0,
\label{conditionbiase}
\end{equation}
is necessary, otherwise the left hand side of Eq. (\ref{GAPim}) vanishes identically. This condition is similar to the existence of threshold voltage in a normal-metal-superconductor junction, where quasi-particle current can flow through the junction when the biased voltage (multiplied by the electric charge) exceeds the BCS energy gap \cite{Tinkham}. When Eq. (\ref{conditionbiase}) is satisfied, the ``tunneling current" from the bath to the exciton BEC gas occurs in the energy region,
\begin{equation}
\Delta_0\le \omega\le \mu_{\rm b}-\mu.
\label{tunnel}
\end{equation}
\par
As mentioned previously, to maintain the non-equilibrium steady state, the leakage of particles into the vacuum must be compensated by the supply of electrons and holes from the bath. In this regard, we note that $\mu_{\rm b}-\mu$ is not sensitive to the decay parameter $\gamma_{\rm v}$ except for $\gamma_{\rm v}/\gamma_{\rm b}\ll 1$. (See the inset in Fig. \ref{fig2}(a).) Thus, when $\gamma_{\rm v}$ increases under a fixed value of the pumping parameter $\gamma_{\rm b}$, the exciton-BEC order parameter $\Delta_0$ must decrease, in order to increase the tunneling flow from the bath by widening the region given in Eq. (\ref{tunnel}). When $\gamma_{\rm v}$ is very large, this compensation mechanism no longer works, leading to the vanishing exciton-BEC, as seen in Fig. \ref{fig2}(a).
\par
\section{Summary}
\par
To summarize, we have investigated a model exciton-BEC gas which is coupled to a bath and a vacuum. Reformulating the BCS-Leggett strong-coupling theory by using the Keldysh Green's function, we have examined how the leakage of electrons and holes into the vacuum, as well as the supply of these particles from the bath, affect the stability of the exciton-BEC. In the non-equilibrium steady state, we showed that the BEC order parameter $\Delta_0$ decreases with increasing the leakage of electrons and holes into the vacuum, to eventually vanish. We also examined the occupied spectral weight function. In the non-equilibrium BEC state, we found that the partial occupation of the upper branch of Bogoliubov single-particle excitations occurs, indicating the depairing of electron-hole pairs. We also pointed out that this phenomenon is deeply related to, not only the decrease of the exciton-BEC order parameter, but also the condition for the stability of the non-equilibrium BEC state. 
\par
In this paper, we dealt with a model electron-hole mixture with a contact interaction at $T_{\rm b}=0$. Extension of this simple treatment to include the realistic long range Coulomb interaction, as well as pairing fluctuations at finite temperatures, is an exciting future problem. Since Fermi condensates out of equilibrium has recently discussed in various systems, not only in an electron-hole gas, but also in an exciton-polariton condensate in a microcavity, as well as in an ultracold Fermi gas, our results would be useful for the development of non-equilibrium Fermi condensates.
\par
\begin{acknowledgements}
We thank M. Yamaguchi, R. Okuyama, D. Inotani, H. Tajima and A. Edelman for useful discussions. R.H. was supported by a Grand-in-Aid for JSPS fellows. This work was supported by KiPAS project in Keio University. YO was also supported by Grant-in-Aid for Scientific research from MEXT and JSPS in Japan (25400418, 15H00840). Work at Argonne National Laboratory is supported by the U.S. Department of Energy, Office of Basic Energy Sciences under contract no. DE-AC02-06CH11357.
\end{acknowledgements}


\end{document}